\documentclass[
reprint, 
superscriptaddress, 
amsmath,
amssymb, 
aps, 
prl,
lengthcheck,
]{revtex4-2}

\usepackage{graphicx}  
\usepackage{dcolumn}   
\usepackage{bm}        
\usepackage{amssymb}   
\usepackage{mathtools} 
\usepackage{physics}
\usepackage{xcolor}    
\usepackage[english]{babel} 
\makeatletter
\providecommand{\l@en}{\l@english}
\makeatother

\usepackage{microtype} 
\usepackage[colorlinks=true,linkcolor=blue,citecolor=blue,urlcolor=blue]{hyperref}

\newif\ifshowurls
\showurlstrue

\begin{document}

\title{Quantized Transport of Gap Solitons in the Harper-Hofstadter Model}

\author{Oliver Ashfield}
    \affiliation{School of Physics and Astronomy, University of Birmingham, Edgbaston, Birmingham, B15 2TT, United Kingdom}

\author{Henry Davenport}
    \affiliation{Blackett Laboratory, Imperial College London, London SW7 2AZ, United Kingdom}

\author{Tom Sheppard}
    \affiliation{School of Physics and Astronomy, University of Birmingham, Edgbaston, Birmingham, B15 2TT, United Kingdom}

\author{David Reid}
    \affiliation{School of Physics and Astronomy, University of Birmingham, Edgbaston, Birmingham, B15 2TT, United Kingdom}

\author{Holly A. J. Middleton-Spencer}
    \affiliation{School of Physics and Astronomy, University of Birmingham, Edgbaston, Birmingham, B15 2TT, United Kingdom}

\author{A. C. Berceanu}
    \affiliation{Extreme Light Infrastructure -- Nuclear Physics (ELI-NP) and Horia Hulubei National Institute for R\&D in Physics and Nuclear Engineering (IFIN-HH), 30 Reactorului Street, 077125 M\u{a}gurele, Romania}

\author{Frank Schindler}
    \affiliation{Blackett Laboratory, Imperial College London, London SW7 2AZ, United Kingdom}

\author{Iacopo Carusotto}
    \affiliation{Pitaevskii BEC Center, INO-CNR and Dipartimento di Fisica,  Università di Trento, Via Sommarive 14, I-38123, Trento, Italy}

\author{Hannah M. Price}
    \affiliation{School of Physics and Astronomy, University of Birmingham, Edgbaston, Birmingham, B15 2TT, United Kingdom}

\date{\today}

\begin{abstract}
    We investigate the dynamics of gap solitons found in the Harper-Hofstadter model with an additional cubic nonlinearity, such as that describing weakly-interacting cold atoms or a Kerr nonlinearity in an optical platform. We find that certain solitons with frequencies deep in the band gap travel with a robust, quantized velocity perpendicular to the applied force and maintain their spatial profile over long timescales. Crucially, this velocity is linearly proportional to the applied force, allowing for arbitrary steering of the nonlinear wavepacket, without breakdown, using only a tunable external force. This quantized transport persists despite a highly non-uniform occupation of the linear Bloch states, and contrasts with solitons near the band edge, which deform and delocalize under the same force. These results point to a robust and highly controllable mechanism for steering nonlinear solitons in experimentally accessible platforms. 
\end{abstract}

\maketitle

{\it Introduction}-- Gap solitons are self-reinforcing, nonlinear wavepackets formed when self-interaction is balanced by the dispersion of an underlying lattice potential.
Studied at length in the context of photonics~\cite{christodoulidesDiscreteSelffocusingNonlinear1988, fleischerObservationDiscreteSolitons2003, fleischerObservationTwodimensionalDiscrete2003, martinDiscreteSolitonsSolitonInduced2004, egorovBrightCavityPolariton2009, chenOpticalSpatialSolitons2012}, and ultracold atoms~\cite{trombettoniDiscreteSolitonsBreathers2001, ostrovskayaMatterWaveGapSolitons2003, eiermannBrightBoseEinsteinGap2004, ankerNonlinearSelfTrappingMatter2005, morschDynamicsBoseEinsteinCondensates2006, cruickshankExperimentalObservationSingle2025}, these localized states can emerge under both attractive and repulsive interactions.
A key feature of gap solitons is that their frequencies lie strictly in the gaps of the single-particle energy bands.
Unlike solitons in free space, gap solitons have no intrinsic motion~\cite{flachMovingLatticeKinks1999}.
They are also difficult--and sometimes impossible--to move smoothly through a lattice using an external field~\cite{kivsharPeierlsNabarroPotentialBarrier1993, morandottiDynamicsDiscreteSolitons1999, naetherPeierlsNabarroEnergySurfaces2011, jenkinsonOnsiteOffsiteBound2015, ablowitzPeierlsNabarroBarrierEffect2021}.

Given the difficulty of moving gap solitons in topologically-trivial band structures, it is natural to ask whether the introduction of band topology can facilitate their transport. In linear systems, the theory of band topology has radically transformed our understanding of transport in lattice models \cite{karplusHallEffectFerromagnetics1954, thoulessQuantizationParticleTransport1983, king-smithTheoryPolarizationCrystalline1993}. 
In two-dimensional (2D) Chern bands (without nonlinearity), transverse transport is closely connected to the physics of the quantum Hall effect~\cite{klitzingNewMethodHighAccuracy1980, thoulessQuantizedHallConductance1982}.
First studied for fermionic systems, the quantum Hall effect occurs when a set of energy bands is fully occupied, with the Hall conductivity quantized according to the total Chern number of those bands.
This Chern number is an example of a topological invariant, an integer-valued quantity which cannot be changed under the continuous variation of model parameters without the closing of a band gap~\cite{niuQuantizedHallConductance1985}.
As a result, the transport properties associated with non-zero Chern numbers are remarkably robust to both perturbations and disorder.

The influence of band topology can also be seen in the semi-classical dynamics of  wavepackets in lattice models.
Specifically, under a weak external force, a wavepacket acquires an anomalous transverse center-of-mass velocity determined by the geometrical Berry curvature averaged over the wavepacket's momentum-space distribution~\cite{changBerryPhaseHyperorbits1995, xiaoBerryPhaseEffects2010, priceMappingBerryCurvature2012, davenport2026compositequantumgeometrysemiclassical}.
The Chern number, corresponding to the integral of the Berry curvature over the Brillouin zone (BZ), is recovered only when the distribution equally samples the entire band.
For bosons, this can occur due to temperature effects or interactions, as exploited experimentally to extract the Chern number in ultracold atoms~\cite{aidelsburgerMeasuringChernNumber2015}.
A related mechanism occurs in driven-dissipative photonic lattices, when the loss rate exceeds the bandwidth (while remaining below the band gap) such that the steady-state response becomes proportional to the Chern number~\cite{ozawa2018steady, chenier2026quantized}.

In these linear systems, however, transport is limited by wavepacket dispersion.
As it travels, a wavepacket generally spreads out in real space. 
This means that, while its center-of-mass position remains traceable, the information it carries (i.e. matter or optical signals) becomes increasingly difficult to resolve.
Stationary gap solitons in non-topological band models do not disperse, by definition.
Naturally, one can ask whether the transport of a gap soliton in a Chern band model is influenced by the topology, specifically if quantized motion can occur.

In this letter, we present the robust and highly controllable quantized motion of gap solitons, with frequencies deep in the band gap, in a nonlinear 2D Chern band model.  
In stark contrast to the usual difficulties in moving these states, we find numerically that their motion can be readily tuned using an external force. 
Remarkably, the solitons do not uniformly sample the band(s) and yet, under an applied force, we find numerically that their velocity is quantized and dependent on the Chern numbers (to very high precision).
We also confirm that gap solitons with frequencies close to the band edge travel according to their semi-classical predictions, as seen in previous work \cite{marzuolaBulkSolitonDynamics2026}. 
Given recent experimental advancements in realizing both topological models~\cite{aidelsburgerRealizationHofstadterHamiltonian2013, miyakeRealizingHarperHamiltonian2013, jotzuExperimentalRealizationTopological2014, rechtsmanPhotonicFloquetTopological2013, roushanChiralGroundstateCurrents2017, owensQuarterfluxHofstadterLattice2018, linn_topological_2026} and 2D gap solitons~\cite{strecker2003bright, cruickshankExperimentalObservationSingle2025, fleischerObservationTwodimensionalDiscrete2003, cerda-mendezExcitonPolaritonGapSolitons2013}, these results open up a highly flexible avenue for the coherent control of macroscopic nonlinear waves.

{\it Model}-- The Harper-Hofstadter model describes a particle that hops on a 2D square lattice with a perpendicular magnetic field $\bm{B} = B \hat{\bm{z}}$~\cite{harperSingleBandMotion1955, hofstadterEnergyLevelsWave1976}.
In the Landau gauge with $\bm{A} = Bx\hat{\bm{y}}$, the linear Hamiltonian (hereafter $e\!=\!\hbar\!=\!1$) is
\begin{align}
	(\mathcal{H}\Psi)_{m, n} =&  -J \big( \Psi_{m+1,n} +\Psi_{m-1,n}
	\\& +e^{-i \phi m} \Psi_{m,n+1} + e^{i \phi m} \Psi_{m, n-1}\big), \label{eq:hh_hamiltonian}\nonumber
\end{align}
where $m, n \in \mathbb{Z}$ are lattice coordinates 
such that $(x,y) \equiv (am,an)$ for a lattice spacing $a$. The hopping strength is~$J$.
We have $\phi=2\pi\alpha $ where $\alpha$ is the number of magnetic flux quanta per plaquette of the lattice.
For a rational flux $\alpha=p/q$ where $p,q \in \mathbb{Z}$ are coprime, the magnetic unit cell has $q$  sites. 
Therefore, there are $q$ bands each of which generally has a non-zero Chern number $C_i$, where $i$ is the band index~\cite{thoulessQuantizedHallConductance1982, avronHomotopyQuantizationCondensed1983, niuQuantizedHallConductance1985, kohmotoTopologicalInvariantQuantization1985, hatsugaiChernNumberEdge1993}.
The eigenstates of $\mathcal{H}$ are the linear Bloch states $\Phi^{\boldsymbol{k}, i}_{m, n}$ labelled by band index $i$ and momentum $\boldsymbol{k}$. These obey $(\mathcal{H}\Phi^{\boldsymbol{k}, i})_{m, n} = \varepsilon_i(\boldsymbol{k})\Phi^{\boldsymbol{k}, i}_{m, n}$ with energy $\varepsilon_i(\boldsymbol{k})$.
Experimentally, this model has been realized using artificial gauge fields in both ultracold atoms ~\cite{aidelsburgerRealizationHofstadterHamiltonian2013, miyakeRealizingHarperHamiltonian2013, aidelsburgerMeasuringChernNumber2015,tai2017microscopy} and photonic platforms ~\cite{hafezi2013imaging,mittal2014topologically,roushan_chiral_2017, linn_topological_2026}.

To study gap solitons in this model, a cubic nonlinearity is added to the Hamiltonian.
The resulting equation is a discrete nonlinear Schrodinger equation given by~\cite{powell2010interacting,powell2011bogoliubov,ozawa2015momentum,fu2022dynamical, reid2026phases}
\begin{eqnarray}
	i \frac{\partial \Psi_{m,n}}{\partial t} &=& (\mathcal{H}\Psi)_{m, n}  + Jg |\Psi_{m,n}|^2 \Psi_{m,n}. \label{eq:hh_dnltdse}
\end{eqnarray}
We search for localized, stationary states of Eq (\ref{eq:hh_dnltdse}) by using the ansatz $\Psi_{m,n}(t) = e^{-i \mu t} \psi_{m,n}$ where $\mu$ is the soliton frequency.
The results in this letter are presented for attractive interactions ($g=-1$), however, we find similar results for both repulsive interactions and saturable nonlinearities.

\begin{figure}
	\includegraphics[width=\columnwidth]{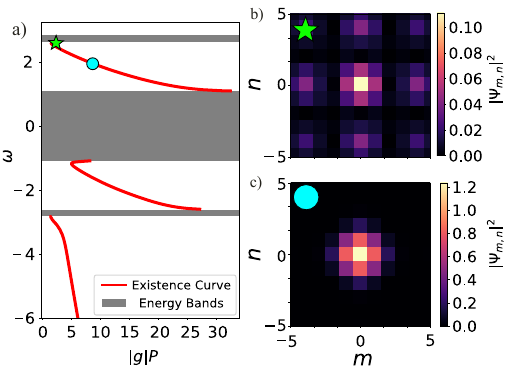}
	\caption{\label{fig:soliton_existence_curve}(a) The existence curve of solitons found in the Harper-Hofstadter model at $\alpha=1/4$ for attractive interactions ($g=-1$).
		The single-particle energy bands (\textit{gray shaded regions}) are shown along with the frequency of the solitons (\textit{red solid line}) as a function of the scaled power $|g|P$.
		It can be seen that the frequency of the solitons is restricted to the energy band gaps.
		Note also that there is a minimum power requirement in a given gap for solitons to form; a known feature of 2D gap solitons~\cite{ostrovskayaPhotonicCrystalsMatter2004, yangNonlinearWavesIntegrable2010}.
		(b) The density profile of a soliton with $\omega=2.605$ and $P=2.33$, at a frequency close to the linear band edge (band-edge). (c) The density profile of a soliton with $\omega=2$ and $P=7.95$ at a frequency deeper in the band gap (mid-gap).
	}
\end{figure}
{\it Solitons in the Harper-Hofstadter Model} -- Eq (\ref{eq:hh_dnltdse}) was solved numerically using a Newton-conjugate-gradient method~\cite{yangNewtonconjugategradientMethodsSolitary2009} to find gap soliton solutions at a given $\omega=\mu/J$.
Given an initial condition close to a soliton solution, the solver converges on a solution with a power $P= \sum_{m,n} |\psi_{m,n}|^2$.
The initial state we use is a Kronecker delta function projected into a single band~\cite{supp}.
In Fig.~\ref{fig:soliton_existence_curve}(a) the frequency $\omega$ of the solitons is plotted against the scaled power $|g|P$ for $\alpha=1/4$.
It can be seen that the frequency of the solitons is strictly within the band gaps, matching previous work on gap solitons found in 2D Chern band models~\cite{liTopologicalBulkSolitons2022, marzuolaBulkSolitonDynamics2026}.

Fig.~\ref{fig:soliton_existence_curve}(b) shows a low power soliton with a frequency close to the linear band edge (a band-edge soliton). We find that the spatial profile of band-edge solitons is more  delocalised and has an underlying periodic structure.
This matches standard 2D gap solitons in non-topological band models, where the wavefunction is well described as a localized envelope modulating the underlying Bloch states of the band minima~\cite{desterkeEnvelopefunctionApproachElectrodynamics1988}.
Fig.~\ref{fig:soliton_existence_curve}(c) shows a higher power with a frequency deeper within the band gap (a mid-gap soliton).
As $P$ increases, mid-gap solitons tend to localize more and lose any underlying periodic structure.

\begin{figure}
	\includegraphics[width=\columnwidth]{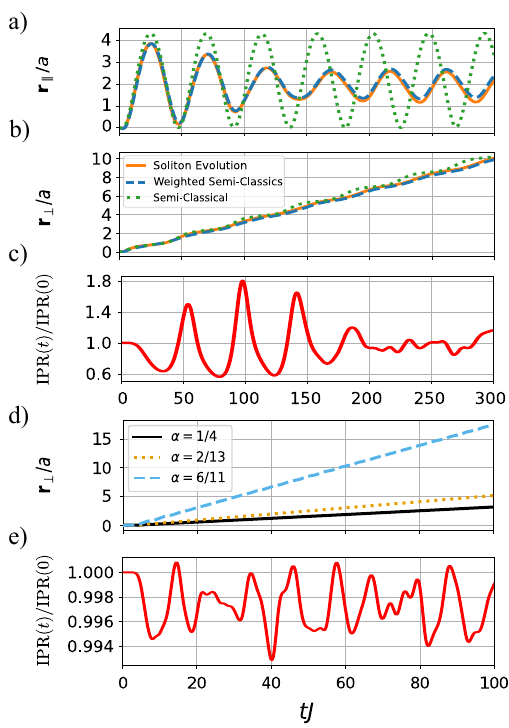}
	\caption{\label{fig:soliton_motion}
		The centre-of-mass (C.O.M) dynamics for various gap solitons.
		All solitons are evolved under the external force 
        with $F=0.05J/a$, $\theta=\pi/4$, $\tau=1/J$, and $t_0=3\tau$.
		(a-b) The parallel ($\mathbf{r}_{\parallel}$) and perpendicular ($\mathbf{r}_{\perp}$) motion, relative to the direction of the force, of a band-edge $\alpha=1/4$ soliton ($\omega=2.605$ and $P=2.33$) (\textit{orange solid line}) along with the weighted (\textit{blue dashed line}) and un-weighted (\textit{green dotted line}) semi-classical motion.
		(c) The normalized inverse participation ratio. (d) The perpendicular C.O.M motion of  mid-gap solitons with: $\alpha=1/4$ (\textit{solid black line}) at $\omega=2.00$ and $P=7.95$ bifurcating from the top band with $C_4 = 1$; $\alpha=2/13$ (\textit{dotted gold line}) at $\omega=2.5$ and $P=11.81$ bifurcating from the two top bands with $C_{13}=-6$ and $C_{12}=7$; and $\alpha=6/11$ (\textit{dashed blue line}) at $\omega=-0.14$ and $P=3.12$ bifurcating from the middle band with $C_6 = 2$.
		The parallel motion in all three cases is negligible.
		(e) The normalized IPR of the $\alpha=1/4$ soliton from (d) undergoing quantized motion.}
\end{figure}

{\it Band-Edge Soliton Dynamics}-- We have studied the dynamics of these gap solitons under a weak external force. We find that ramping the force prevents oscillations in the soliton density which disturb the desired nonlinear dynamics. We therefore apply the force $\boldsymbol{F} = F(t) \hat{\boldsymbol{r}}_{\theta}$ where the force is applied at angle $\theta$ relative to the lattice $\hat{\boldsymbol{r}}_{\theta}\equiv \cos(\theta) \hat{\mathbf{x}} + \sin(\theta) \hat{\mathbf{y}}$.
The magnitude of the force is $F(t)\equiv\frac{F}{2} \left[\tanh\left(\frac{t-t_0}{\tau}\right) + 1 \right]$ where $\tau$ is the ramp timescale, $t_0$ is the offset of ramp. We keep $|F(t)|a \ll E_{BG}$, where $E_{BG}$ is the size of the band gap, to avoid Landau-Zener tunneling~\cite{cristiani_experimental_2002, takahashiLandauZenerTunnelingProblem2017}.
It is known that the presence of nonlinearity can enhance this tunneling~\cite{wuNonlinearLandauZenerTunneling2000, jona-lasinioAsymmetricLandauZenerTunneling2003}.

Fig.~\ref{fig:soliton_motion}(a) and (b) shows the resulting dynamics for the band-edge soliton depicted in Fig.~\ref{fig:soliton_existence_curve}(b).
The instantaneous velocity is well-approximated by the adiabatic density-weighted semi-classical prediction for the linear system (i.e. Eq.~\eqref{eq:hh_dnltdse} with $g=0$) within a single-band approximation. This idea has previously been explored in a topological Lieb lattice~\cite{marzuolaBulkSolitonDynamics2026}.
Specifically, for a semi-classical wavepacket with center-of-mass position $\langle \mathbf{r} \rangle$, this is given by
\begin{equation}
	\langle{\dot{\mathbf{r}}(t) \rangle}= \sum_{\mathbf{k}} |\tilde{\psi}_i(\mathbf{k}, t)|^2 \Biggl(\nabla_{\mathbf{k}} \varepsilon_i(\mathbf{k}) - \mathbf{F}\times \mathbf{\Omega}_i (\mathbf{k)}\Biggr), \label{eq:semi}
\end{equation}
with the Berry curvature $\mathbf{\Omega}_i (\mathbf{k}) = {\Omega}_i (\mathbf{k})\hat{\bm{z}}$ at momentum~$\boldsymbol{k}$~\cite{changBerryPhaseHyperorbits1995, xiaoBerryPhaseEffects2010, davenport2026compositequantumgeometrysemiclassical}.
Nonlinear effects are included within the time-dependent overlap with the Bloch states i.e. $\tilde{\psi}_i(\mathbf{k}, t) \equiv P^{-1}\sum_{m, n}\left(\Phi_{m, n}^{\boldsymbol{k}, i}\right)^*\Psi_{m, n}(t)$ as they detail the evolution of the momentum density. Since the interactions are density-dependent, weighting by these overlaps also accounts for the evolution of the nonlinear term  ~\cite{royWavepacketDynamicsChernband2015}.
For Fig.~\ref{fig:soliton_motion}(a) and (b), only the $i=4$ band overlap was included as the soliton has negligible overlap with other bands.

The semi-classical equation of motion [Eq.~\eqref{eq:semi}] gives rise to two key dynamical phenomena: a transverse ``anomalous" drift and Bloch oscillations~\cite{witthautBlochOscillationsTwodimensional2004, priceMappingBerryCurvature2012}.
The anomalous drift originates from the Berry curvature [i.e. the second term in Eq.~\eqref{eq:semi}] and is always perpendicular to the applied force.
For this specific band, the Berry curvature has the same sign across the BZ, leading to a continual increase in the transverse center-of-mass position; this dominates the response in Fig.~\ref{fig:soliton_motion}(b).
However, importantly, the corresponding anomalous velocity is not constant as the Berry curvature is not uniform.
Secondly, the Bloch oscillations originate from the group velocity [i.e. the first term in Eq.~\eqref{eq:semi}].
As the force is not aligned with the lattice axes in Fig.~\ref{fig:soliton_motion}, the Bloch oscillation corresponds to a complicated trajectory in real space, contributing to the motion along both the perpendicular and parallel directions~\cite{priceMappingBerryCurvature2012}.

For contrast, in Fig.~\ref{fig:soliton_motion}(a) and (b) we also show the unweighted semi-classical prediction for $\mathbf{k}=(\pi/4a, \pi/4a)$ (i.e. the band minimum in momentum space).
We see that this does not predict the soliton evolution as accurately; this is because the band-edge solitons deform as they evolve.
This can also be seen in the normalized inverse-participation ratio, $\text{IPR}(t) = \sum_{m,n}|\psi_{m,n}(t)|^4/P^2$ [Fig.~\ref{fig:soliton_motion}(c)].
The IPR measures the localization of the soliton, with a lower IPR corresponding to a more delocalized state.
Fig.~\ref{fig:soliton_motion}(c) shows that the $\text{IPR}$ of the band-edge soliton fluctuates significantly during the evolution.
This can be understood as the applied force driving $\langle {\bf k} \rangle$ into a region of the BZ where the effective mass changes sign (i.e.\ from near a band minimum to a band maximum).
This causes the gap soliton to lose support and become unstable~\cite{supp}. Hence, the band-edge solitons are not robust under an applied force.

{\it Mid-Gap Soliton Dynamics}-- Under the same applied force, we find that mid-gap solitons can travel with a robust, quantized velocity perpendicular to the force.
As seen in Fig.~\ref{fig:soliton_motion}(d), the perpendicular motion (\textit{black solid line}) of the $\alpha=1/4$ mid-gap soliton depicted in Fig.~\ref{fig:soliton_existence_curve}(c) shows no Bloch oscillations but remarkably does show a quantized transverse velocity.
The parallel motion is not plotted as it is negligible compared to the transverse motion.
We find numerically that the center-of-mass velocity $\mathbf{v}(t) \equiv \langle \dot{\mathbf{r} }(t) \rangle$ is well-fitted by the expression
\begin{equation}
	\mathbf{v}(t) = \frac{q \nu}{2 \pi} \hat{\mathbf{z}} \times \mathbf{F}(t) \label{eq:linear_quantized_motion_multiple_bands} \end{equation}
where $\nu$ is a constant.
For the $\alpha=1/4$ mid-gap soliton, we find $\nu=1.00208(6)$. Moreover, we find that solitons undergoing this quantized motion are robust to disorder in the lattice ~\cite{supp}. Importantly, we note that this result does not depend on the angle of the applied force with respect to the underlying lattice.

We find quantized motion for mid-gap solitons bifurcating from the top bands of $\alpha=1/q$ band structures with $q>4$; these all give $\nu=1$ to high precision. This exactly coincides with the Chern number of the top bands when $\alpha=1/q$.
However, it should be noted that mid-gap solitons with quantized motion are not guaranteed in every band gap.
Generally, we find that solitons bifurcating from bands with a band flatness (i.e. a ratio of the band gap $E_{BG}$ to the bandwidth $E_{BW}$) that is large are the most reliable candidates for quantized motion.
However, the role of flatness is not clear as discussed in the supplementary material~\cite{supp}.

Since solitons bifurcating from bands with Chern number $C_i=1$ exhibit a quantized velocity with $\nu
=1$, we investigate bands with higher Chern numbers to test whether this relationship holds more generally.
We begin with an example of a band with Chern number $C_i=2$. We obtain a mid-gap soliton bifurcating from the middle band of $\alpha=6/11$ [Fig.~\ref{fig:soliton_motion}(d) {\it blue dashed line}]. 
This moves with a velocity well fitted by $\nu=1.9997(3)$ i.e. identical to the Chern number within error. 
Secondly, we study $\alpha=2/13$, where the top two bands are separated by a small but non-zero energy gap. 
Unlike the previous solitons discussed, which had most of their weight in a single band, the converged soliton has significant overlap with both bands. The resulting soliton velocity ({\it gold dotted line}) is now given by the fractional $\nu = 0.49995(8)$. 
This same result holds for solitons at $\alpha=2/q$ that overlap strongly with the top two (nearly degenerate) bands. 
Similarly, $\nu=1/3$ for $\alpha=3/q$ when the soliton overlaps strongly with the top three (nearly degenerate) bands.

The mid-gap soliton's quantized motion appears to be related to the topology of the underlying Harper-Hofstader model bands.
For solitons that bifurcate from a single band, the numerically observed values of  $\nu$ can be reproduced from Eq.~\eqref{eq:semi} by assuming the band from which the soliton bifurcates is uniformly filled (i.e. $|\tilde{\psi}_i(\mathbf{k}, t)|^2 = 1/L$). For solitons bifurcating from multiple bands (i.e. for $\alpha = 2/q$) the observed $\nu$ can be reproduced by assuming that the bands are equally weighted in the soliton wave function, with each band uniformly occupied. This is because, if multiple bands are uniformly occupied, $\nu$ is given by $C_{\mathrm{eff}}=\sum_i \rho_i C_i$ with $\rho_i = \sum_{\mathbf{k}} |\tilde{\psi}_i(\mathbf{k}, t=0)|^2$ being the total overlap of the wavefunction with the given band. These predictions match the observed results as it can be easily calculated that: for $\alpha=1/q$, $C_q=1$; for $\alpha=6/11$, $C_6=2$; and for $\alpha=2/7$, $C_7=4$ and $C_6=-3$ giving $C_{\mathrm{eff}}=1/2$ when the two bands are occupied evenly with $\rho_7=\rho_6=1/2$.

However, it is not true that the bands are uniformly occupied and contribute with equal weight to the soliton wave function. Firstly, the weight $|\tilde{\psi}_i(\mathbf{k}, t)|^2$ can vanish over much of the BZ even as the soliton exhibits a quantised velocity set by the Chern number of the band. Secondly, mid-gap solitons exhibiting quantized motion often overlap with more bands than those included in the above linear prediction, and these overlaps vary strongly with the power even while the quantization remains fixed. The above assumption of equal filling of bands within a nearly-degenerate subset is also not true; for $\alpha=2/7$, we find $\rho_7\approx0.53$ and $\rho_6\approx0.40$, which predicts $C_{\mathrm{eff}}=0.92$ instead of the observed $\nu=0.5$. This shows that the center-of-mass velocity cannot straightforwardly be interpreted as a linear quantum Hall response~\cite{supp}.

\begin{figure}
	\includegraphics[width=\columnwidth]{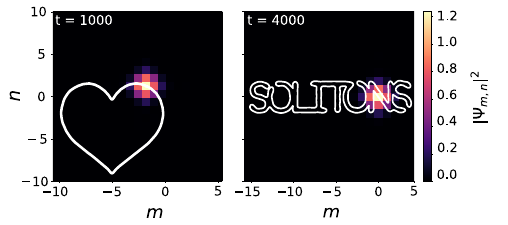}
	\caption{\label{fig:soliton_stability}
		An $\alpha\!
			=\!1/4$ mid-gap soliton ($P\!=\!7.95, \, \omega\!=\!2$) tracing out an arbitrary, continuous curves in real-space using only a time-dependent external force $\mathbf{F}(t)$ where $|\mathbf{F}(t)|a\ll E_{\mathrm{BG}}$.}
\end{figure}

{\it Controllable Motion}--Mid-gap solitons undergoing quantized motion also maintain their profile as they evolve, resulting in very stable dynamics. 
At $\alpha=1/4$, the IPR of the mid-gap soliton [Fig.~\ref{fig:soliton_motion}(e)] shows that it remains highly localized, unlike the band-edge soliton [Fig.~\ref{fig:soliton_motion}(c)].
Similar behaviour is observed for all solitons we have found undergoing quantized motion.
Moreover, it is possible to achieve near-complete control over the soliton motion without any dispersion. In Fig.~\ref{fig:soliton_stability} we show that the soliton can trace an arbitrary continuous curve. 
Note that this requires again that the applied force is small compared to the band gap.

This connection to linear Chern numbers is reminiscent of the topological quantization observed in the nonlinear Thouless pumping of gap solitons in time-dependent, Chern band models~\cite{jurgensenQuantizedNonlinearThouless2021, jurgensenChernNumberGoverns2022a, fuNonlinearThoulessPumping2022, mostaanQuantizedTopologicalPumping2022, citroThoulessPumpingTopology2023, jurgensenQuantizedFractionalThouless2023, bestlerQuantizedNonlinearKink2025, wuTopologicalInvariantsNonlinear2026, bohmQuantumTheoryFractional2026}. However, in nonlinear Thouless pumping, the Hamiltonian is varied cyclically in time. The soliton then adiabatically follows the instantaneous eigenstates of the Hamiltonian with a quantized displacement only after a pump cycle.
Here, by contrast, the lattice is time independent, and the quantization appears directly in the soliton velocity, which can be controlled continuously through the force, directed at any angle with respect to the lattice.

\begin{figure}
\includegraphics[width=\columnwidth]{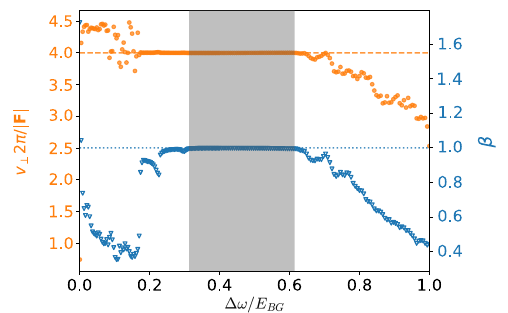}
	\caption{\label{fig:soliton_breakdown} Existence of quantized motion for $\alpha=1/4$ gap solitons bifurcating from the top $i\!=\!4$ single-particle band.
    A dual plot showing the fitted velocity $v_{\perp}$ and $\beta$ against the difference $\Delta\omega$ of the soliton frequency from the band minimum as a fraction of the band gap $E_{\text{BG}}$. 
    The error bars are each point are omitted due to them being too small.
    Dashed lines indicate the values of $v_{\perp}$ and $\beta$ corresponding to ideal quantized motion for these parameters.
    The shaded grey region is added by eye to indicate the range of solitons undergoing quantized motion.
    All solitons were found for $g=-1$ and evolved under a force with $F=0.05J/a$, $\tau=1/J$, $t_0=3 \tau$ and $\theta=\pi/4$.
    The evolution was run until $t_{\text{final}}=300/J$ to capture long time dynamics of each soliton.
	}
\end{figure}

{\it Regimes of Quantized Motion}-- Importantly, the quantized motion of mid-gap solitons is not a fine-tuned phenomenon, but is observed for a wide-range of soliton frequencies $\omega$ within the band-gap.
As discussed above, the motion is characterized by a quantized velocity and a soliton profile which remains localized as it travels under an applied force.
We therefore identify the regime of quantized motion by numerically calculating the transverse velocity ($v_{\perp}$) as well as the long-time average of the normalized IPR ($\beta$). Solitons which remain localized throughout their evolution have $\beta=1$. See End Matter for details of how these are defined. 
Fig.~\ref{fig:soliton_breakdown} shows a dual plot of $v_{\perp}$ and $\beta$ for $\alpha=1/4$ gap solitons bifurcating from the top band.
These are plotted against the depth of the soliton frequency into the band gap $\Delta \omega/E_{\mathrm{BG}} = (E_{t}-\omega)/(E_{t}-E_{b})$ where $E_t$ is the energy minimum of the top band and $E_b$ is the maximum of lower band.
The shaded region highlights the large portion of the band gap where solitons undergo quantized motion.
For small values of $\Delta \omega$ we see the breakdown of band-edge solitons with frequencies close to the band minimum, as discussed previously.
On the other hand, large values of $\Delta \omega$ are solitons close to the band maximum of the lower single-particle band ($i=3$) with breakdown similar to the band-edge case.
Similar results have been found for the other solitons analyzed in this paper.
This shows that robust quantized motion can be found without fine-tuning of parameters.

{\it Conclusions} -- We have demonstrated that gap solitons in the nonlinear Harper-Hofstadter model can exhibit robust, quantized motion when driven by an external force.
This transport is characterized by a velocity proportional to the applied force which persists across all numerical timescales we ran and appears to be given by the single-particle Chern number.
Given that gap solitons have been discovered in other nonlinear topological models~\cite{liTopologicalBulkSolitons2022, schindlerNonlinearBreathersCrystalline2025, marzuolaBulkSolitonDynamics2026}, it is natural to investigate whether such controllable motion of gap solitons, a previously difficult task in 2D lattices, can be seen in other models.
These findings provide a blueprint for the arbitrary steering of solitons, enabling stable transport of optical or matter waves without spreading or breakdown.

{\it Acknowledgements}--We thank Mike Gunn, Oded Zilberberg, Markus Bestler, Nigel Cooper, Joe Bhaseen, and Wladimir Benalcazar for helpful discussions.
This work is supported by the Royal Society via grants URF\textbackslash R\textbackslash221004 and RGF\textbackslash{}EA\textbackslash{}180121 and by the Engineering and Physical Sciences Research Council [grant numbers EP/W016141/1, EP/Y01510X/1 and UKRI2226].
H.~D.~acknowledges support from the Engineering and Physical Sciences Research Council (grant number EP/W524323/1). H.~D.~and F.~S.~were supported by a UKRI Future Leaders Fellowship MR/Y017331/1.
A.~C.~B.~acknowledges support from the Extreme Light Infrastructure Nuclear Physics Phase II project, co-financed by the Romanian Government and the European Union through the European Regional Development Fund and the Competitiveness Operational Programme (No.~1/07.07.2016, COP, ID~1334) and ELI-RO/DEZ/2023\_001, funded by the Romanian Ministry of Education and Research.
The computations described in this paper were performed using the University of Birmingham’s Bluebear HPC service, which provides a High Performance Computing service to the University’s research community.

{\it Data Availability}--The data of the numerical results presented in this letter are available from the authors upon reasonable request.
\\

\ifshowurls\else
	\renewcommand{\urlprefix}{}%
	\renewcommand{\url}[1]{}%
\fi
\bibliographystyle{apsrev4-2}
\bibliography{references_zotero, references_collaborators}

\onecolumngrid
\vspace{1.5em}
\begin{center}
    {\large\bfseries End Matter}
\end{center}
\vspace{1em}
\twocolumngrid

\appendix

\setcounter{equation}{0}
\renewcommand{\theequation}{A\arabic{equation}}

\textit{Definition of Breakdown Parameters}--Taking Eq. \eqref{eq:linear_quantized_motion_multiple_bands} along with the ramped force $\mathbf{F}(t)$ we use in the main text
\begin{equation}
    \mathbf{F}(t)=\frac{F}{2} \left[\tanh \left(\frac{t-t_0}{\tau}\right) +1\right] \hat{\mathbf{r}}_{\theta}
\end{equation}
we can integrate the velocity to find the position evolution of a soliton undergoing ideal quantized motion. As this transport is only perpendicular to the applied force, we only need to solve for the position in this perpendicular direction. Integrating Eq. \ref{eq:linear_quantized_motion_multiple_bands} with this force ramp gives
\begin{equation}
    r_{\perp}(t)=\frac{v_{\perp}}{2} \left\{\tau \ln \left[\frac{\cosh \left(\frac{t-t_0}{\tau}\right)}{\cosh\left(\frac{t_0}{\tau}\right)}\right] + t \right\} \label{aeq:position_fit}
\end{equation}
where $v_{\perp}=q\nu F/2\pi$ for ideal quantized motion and we choose $r_{\perp}(0)=0$. We then use $v_{\perp}$ as a free parameter and fit the soliton motion to Eq. \eqref{aeq:position_fit}. Fig. \ref{fig:soliton_stability} plots this fitted value scaled by the force where the dashed line shows the ideal value. This allows us to identify which solitons in the band gap have a center-of-mass motion which matches the quantized motion. However, this does not contain any information about whether the soliton remains localized during its evolution.

The inverse participation ratio (IPR) measures the soliton localization during its evolution and can be used to determine whether it remains localized as it travels. Due to the ramped force used in the main text, we want to measure the value of the IPR long after the force has saturated to avoid initial transience. Moreover, the IPR exhibits small oscillations about a value close to, but slightly below, its initial value for solitons undergoing quantized motion.
To account for this, we define $\beta$ as a late-time average of the normalized IPR
\begin{equation}
	\beta = \frac{1}{t_{\text{final}} - t_{\text{long}}} \int_{t_{\text{long}}}^{t_{\text{final}}} \left(\frac{\text{IPR}(t)}{\text{IPR}(0)}\right) dt.
\end{equation}
We used $t_{\text{long}}=0.8t_f$ for our simulations, as we found that this gave us the most consistent average. This definition also accounts for periodic boundary conditions unlike the soliton width which can change as the soliton reaches the edge of the finite lattice. The value of $\beta$ for each soliton evolution is plotted in Fig. \ref{fig:soliton_stability} with a dashed line at $\beta=1$ which corresponds to a wavepacket which never disperses (i.e. $\text{IPR}(t)=\text{IPR}(0))$. The controllable quantized motion discussed in the main text requires not only the correct center-of-mass motion but also that the soliton does not break down as it evolves. These two parameters can therefore determine which solitons in the band gap are valid candidates for controllable motion. 

\end{document}